\documentclass[twoside]{mhd}

\usepackage{siunitx}
\usepackage{amsmath}
\usepackage{amssymb}
\usepackage{graphicx}
\usepackage{subfig}
\usepackage{booktabs}
\usepackage{multirow}
\usepackage{array}
\usepackage{nameref}
\usepackage{microtype}

\mhdhead{55}{1}{1}	

\title{Convection-Caused Symmetry Breaking of Azimuthal Magnetorotational Instability in a Liquid Metal Taylor-Couette Flow}

\author{M. Seilmayer, J. Ogbonna, F. Stefani}

\institute{ Helmholtz-Zentrum Dresden-Rossendorf, Bautzner Landstr. 400, 01328 Dresden, Germany} 

\global\long\def\laplace{\mathop{}\!\mathbin{\bigtriangleup}}

\begin{document}
\maketitle

% ------------------------------ Abstract --------------------------------

\begin{abstract}
We report the results of a liquid metal Taylor-Couette experiment in the Rayleigh-stable regime under the influence of an azimuthal magnetic field. We observe that the resulting azimuthal magnetorotational instability (AMRI) from our experimental setup is significantly influenced by the thermal boundary conditions. Even a minimal radial heat flux leads to a symmetry breaking, which results in the AMRI waves traveling either upwards or downwards. 
We identify the thermal radiation by the central axial current as the heat source responsible for vertical convection in the liquid. 
Preliminary numerical investigations point towards an interaction between AMRI and thermal convection, which supports our experimental findings.
\end{abstract}

% --------------------------- Introduction --------------------------------

\section*{Introduction}

The magnetorotational instability (MRI) is the main candidate to explain outward-directed angular momentum transport and inward-directed mass flow in accretion discs around protostars and black holes. To date, only MRI in azimuthal (AMRI) \cite{Seilmayer2014} and helical (HMRI) \cite{Stefani2009} magnetic field configurations have been experimentally observed, while the unanimous demonstration of the original standard MRI with a purely axial magnetic field is still elusive. In our experiments, the shear flow between the two cylinders with inner and outer radii of $r_\mathrm i = \SI{4}{\centi\meter}$ and $r_\mathrm o = \SI{8}{\centi\meter}$, respectively, is in the hydrodynamically stable regime according to Rayleigh's criterion, with $\Omega_\mathrm{o} / \Omega_\mathrm{i} > \num{0.25}$. The AMRI occurs as a non-axisymmetric wave with azimuthal wavenumber $m = \pm 1$ with a characteristic drift frequency $\omega_\mathrm{AMRI}$ and axial wavenumber $k_\mathrm{AMRI}$.

For the experiment, the shear flow in a Taylor-Couette (TC) setup is exposed to an azimuthal magnetic field $B_\varphi \propto r^{-1}$.
The measured axial velocity component of AMRI $u_\mathrm z = \mathcal{O}(\SI{0.1}{\milli\meter\per\second})$ is inferred from an Ultrasound Doppler Velocimetry (UDV) system. 
For this purpose, two UDV sensors are located at the top of the annulus, co-rotating with the outer cylinder.
In the initial AMRI experiment \cite{Seilmayer2014} with the PROMISE facility, a frame of copper rods on one side of the setup supplied the central axial current by which the required magnetic field was produced. However, the lopsided nature of the frame introduced a relative field inhomogeneity with $m = 1$ symmetry of about $\Delta B_\varphi / B_0 \lesssim \SI{8}{\percent}$. 
The slightly asymmetric field led to a stationary background flow, which adapted to the residual $m = 1$ magnetic field disturbance (see~\cite{Seilmayer2016} for details). 
Moreover the bottom and top lids introduce an additional axial symmetry breaking, which causes the wave to separate into the two possible $m = \{-1,1\}$ modes with some preference for either the top or bottom half of the cylinder (see \cite[Fig.~7]{Seilmayer2016}).

\begin{figure}
\hfill{}
  \subfloat[\label{fig:CrossSection}Cross section]{\includegraphics[width=0.5\columnwidth]{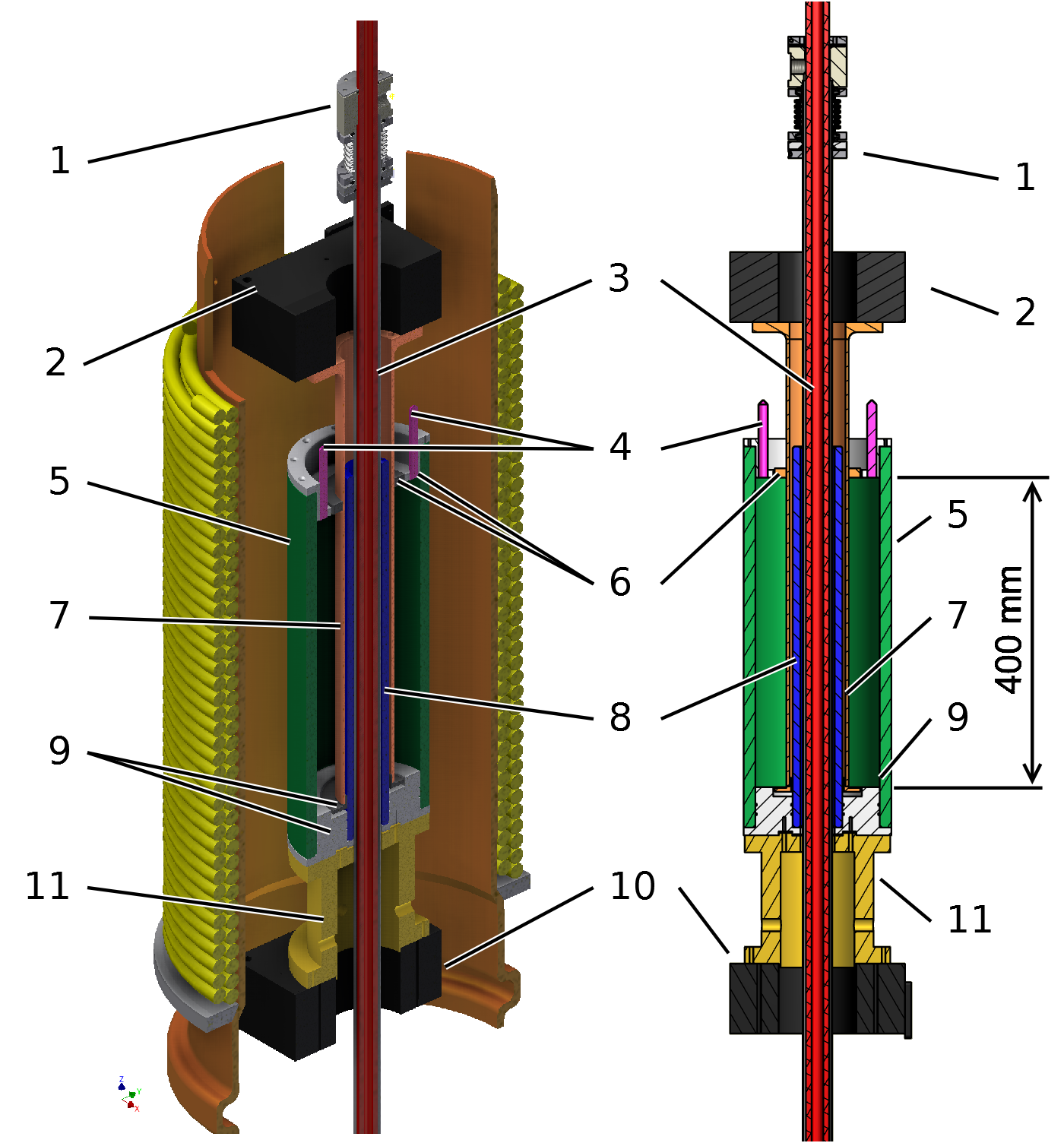}}
\hfill{}
  \subfloat[\label{fig:Experimental-setup-Foto}Current return]{\includegraphics[height=7cm]{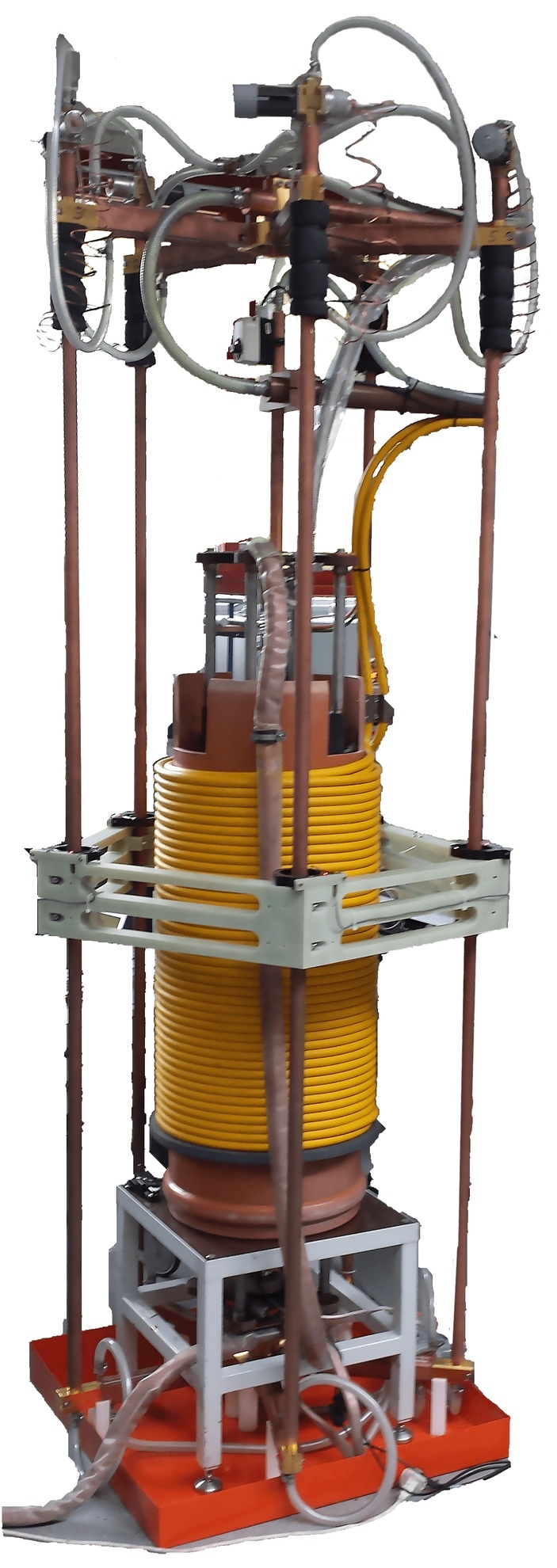}}
\hfill{}
  \caption{\label{fig:Experimental-setup}PROMISE setup. a) Cross-section of the experiment with main dimensions $h=\SI{0.4}{\meter}$, $r_{\mathrm{i}}=\SI{40}{\milli\meter}$ and $r_{\mathrm{o}}=\SI{80}{\milli\meter}$:
(1) Vacuum insulation; (2) Upper motor; (3) Current-carrying copper
rod; (4) UDV sensors; (5) Outer cylinder; (6) Top acrylic glass split
rings; (7) Inner cylinder; (8) Center cylinder; (9) Bottom split rings;
(10) Bottom motor; (11) Interface. b) The improved magnetic
field system consists of five balanced return paths.}

\end{figure}

Since then, several improvements have been implemented to minimize the asymmetry of the magnetic field, with an aim to weaken the non-axisymmetric background flow. 
During the systematic re-investigation of the AMRI, it was now discovered that a weak heat flux originating from the central current (see Fig.~\ref{fig:CrossSection} -- 1) played a crucial role in determining the characteristics and the symmetry of the AMRI. 
It must be mentioned however, that the PROMISE facility was not designed with thermal convection experiments in mind. Hence, the control and measurement of the thermal boundaries in a rotating system like that of PROMISE remains a major challenge. 

\raggedbottom

% ---------------------------- Improvements -------------------------------
\section{Improvements of the facility}

The main improvement of the PROMISE setup with respect to the installation underlying \cite{Seilmayer2014} is the pentagon-shaped current return path ({Fig.~\ref{fig:Experimental-setup-Foto}}), which largely eliminates the $m = 1$ background flow observed in previous experiments \cite{Seilmayer2014, Seilmayer2016}. 
The optimized magnetic field system provides a uniform $B_\varphi$ distribution along $\varphi$ with a strongly reduced field disturbance. 
Measurements prove a relative deviation of $\Delta B_\varphi / B_0 < \SI{0.5}{\percent}$, which is comparable to the theoretical value of \num{1e-4} for the five return path configuration \cite{Seilmayer2016a}. 

Additionally, a misalignment of the inner and outer cylinder axes and the central rod axis impose a weak residual $m = 1 $ modulation of the magnetic field. With the new improvement, the absolute displacement of the fluid by the magnetic field was minimized to a value in the order of \SI{1.5}{\milli\meter}.

\begin{figure}
	\subfloat[\label{fig:Drift}Drift frequencies]{\includegraphics[width=0.49\columnwidth]{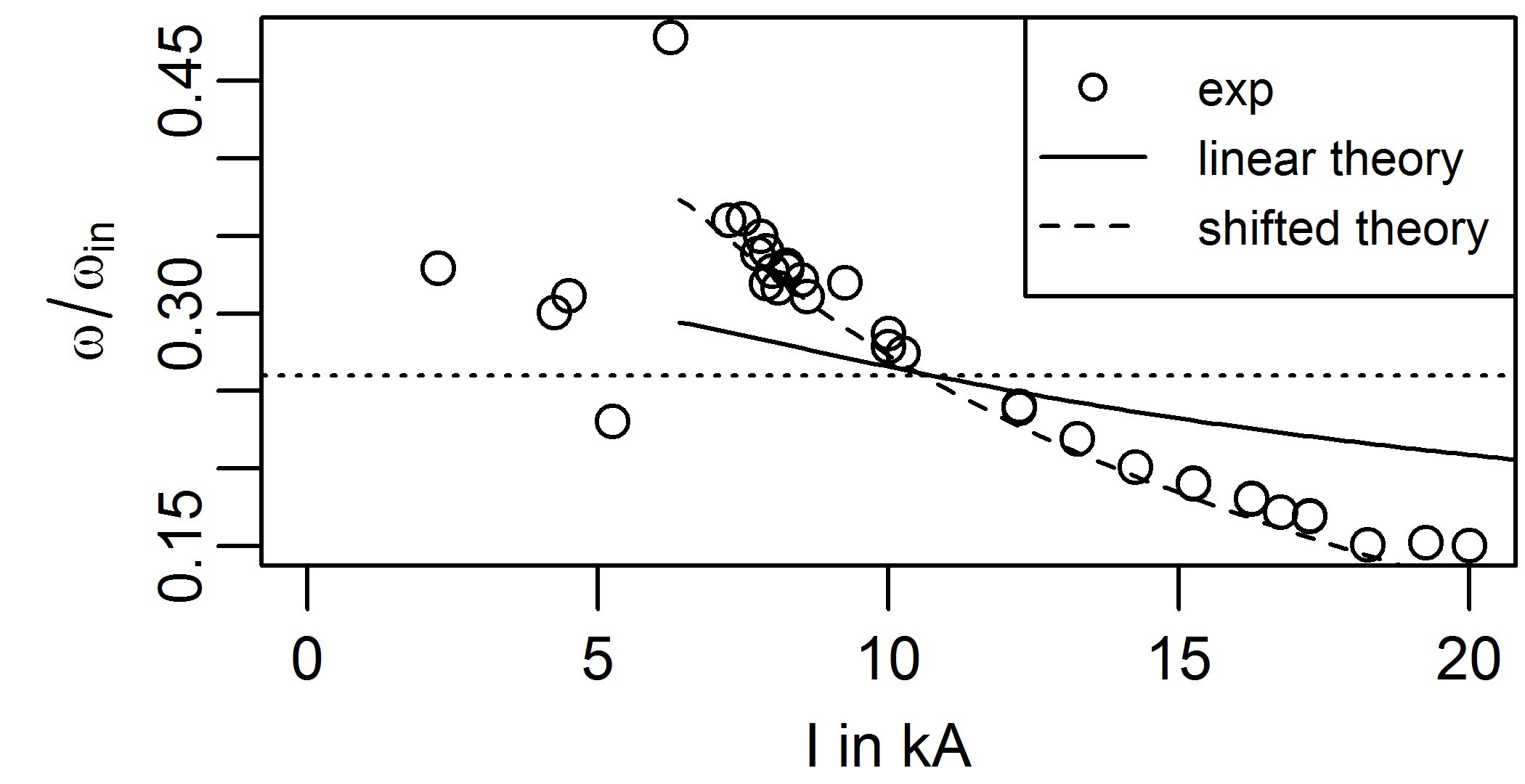}}
		\hfill{} 
	\subfloat[\label{fig:Wavenumber}Wave numbers]{\includegraphics[width=0.49\columnwidth]{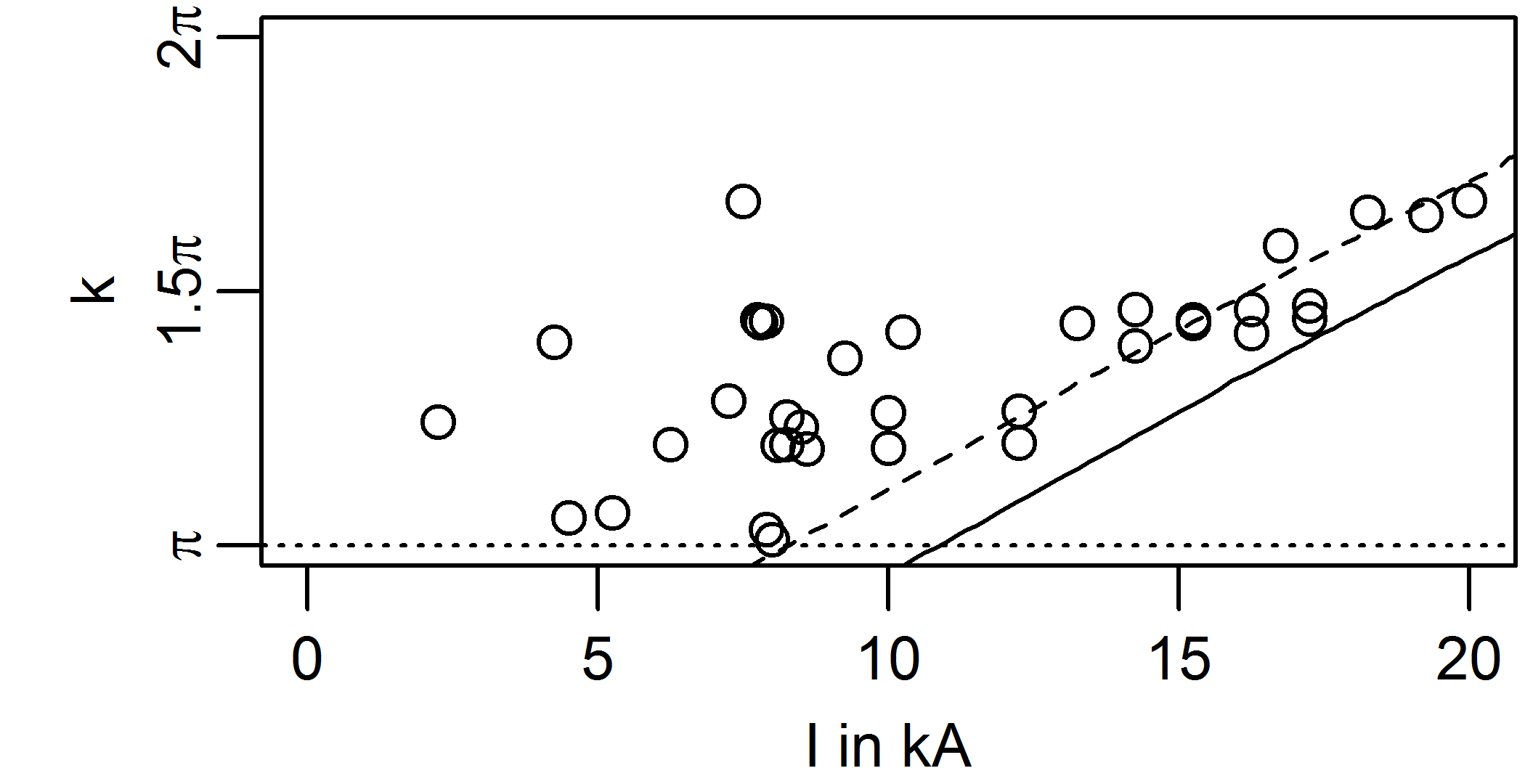}} \\

	\subfloat[\label{PhaseVelo}Phase velocities]{\includegraphics[width=0.49\columnwidth]{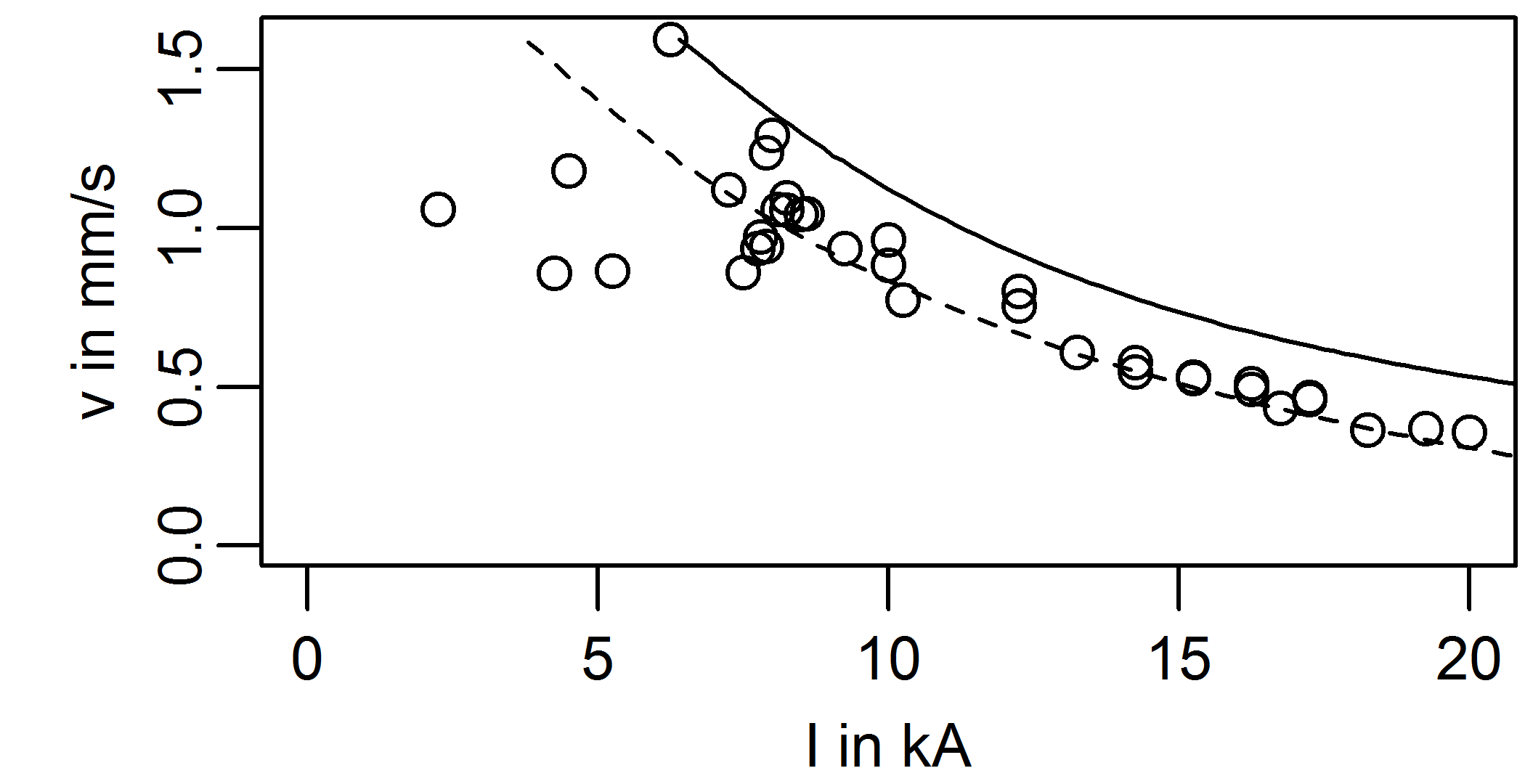}}
		\hfill{}
	\subfloat[\label{fig:Energy}Energies]{\includegraphics[width=0.49\columnwidth]{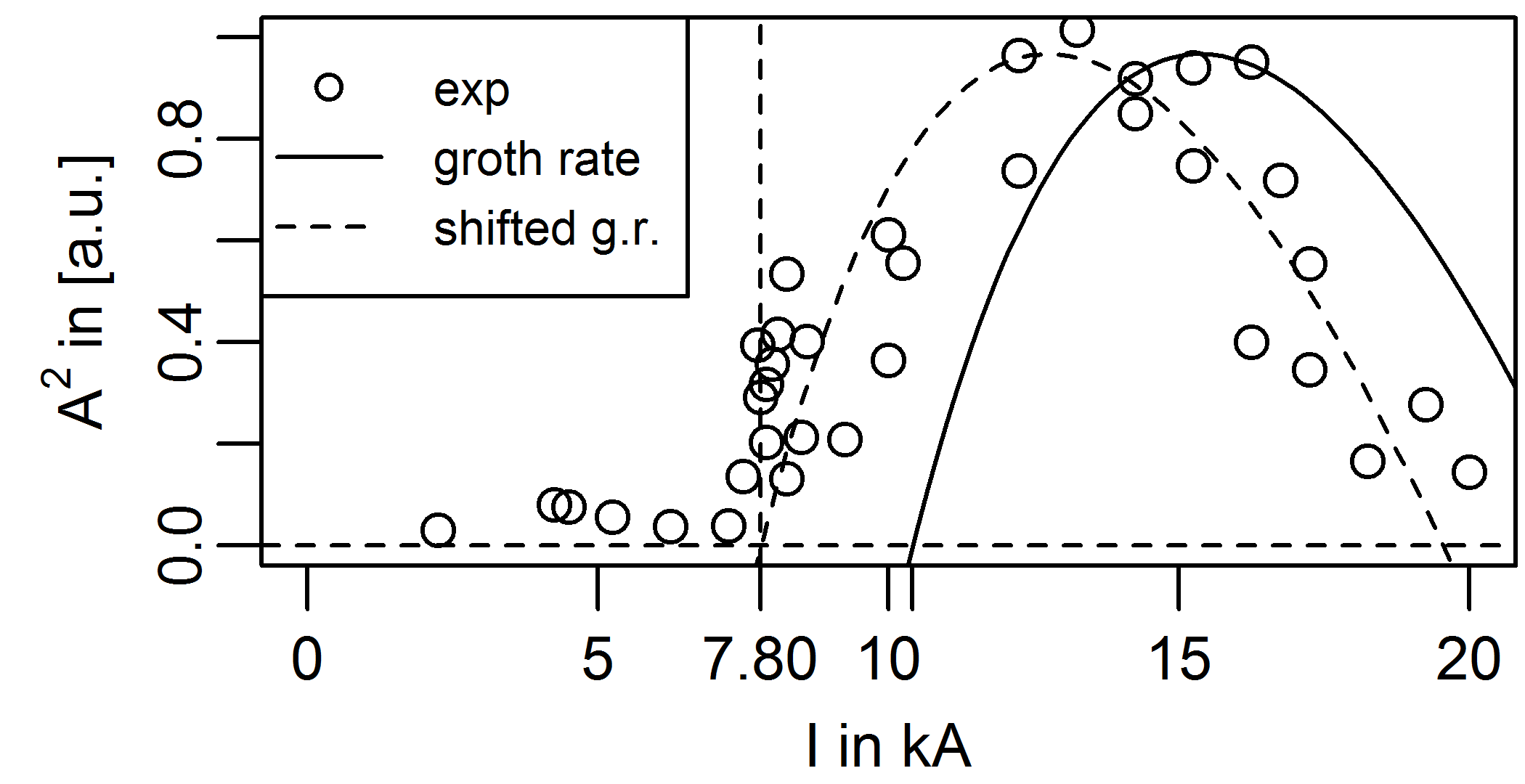}}

	\caption{Experimental results with the improved magnetic field system. 
	a) The frequencies of the significant waves systematically deviate from results of the linear stability analysis (solid line). 
	b) Wave numbers.  
	c) The phase velocities fit in a curve in the AMRI unstable regime. 
	d) Energy contents of the AMRI wave. A scaled and shifted growth rate curve is included for comparison.
	}
	\label{fig:AMRI-results}

\end{figure}

With the improved installation, the AMRI was re-investigated in detail with $\Omega_\mathrm i = 2\pi \cdot \SI{0.05}{\hertz}$ and $\Omega_\mathrm o = 2\pi \cdot \SI{0.013}{\hertz}$. Figure~\ref{fig:AMRI-results} shows the major results of the investigation. 
First, we find a strong and reasonable dependence of all characteristic qualities, i.\,e. drift rate $\omega$, wave number $k$, phase velocity $v$ and energy content $A^2$, on the Hartmann number $\mathit{Ha} = B_{\varphi}(r_{\mathrm{i}})r_{\mathrm{i}}\sqrt{\sigma / (\rho\nu)} \approx \SI{7.77e-3}{\per\ampere}\cdot I$. 
It is also remarkable that the variations within each characteristic quality were reduced, so that several values fit in a smooth line. However, Figure~\ref{fig:AMRI-results} also shows that the characteristic frequencies of the AMRI wave deviate systematically from linear theory (solid line adapted from \cite{Hollerbach2010}). 
A closer look at the drift rate given in Fig.~\ref{fig:Drift} indicates a dependence of the deviation $\Delta\omega$ on the central current, i.\,e. $\Delta\omega \propto I$. 
From the energy content (Fig.~\ref{fig:Energy}), we observe that the critical current $I_\mathrm{crit} \approx \SI{7.8}{\kilo\ampere}$ ($\mathit{Ha} \approx 60.6$) is lower than the prediction from linear theory, $I_\mathrm{crit,th} \approx \SI{10.4}{\kilo\ampere}$ \cite{Rudiger2018a, Hollerbach2010}.
Furthermore, the main flow feature is an axial symmetry breaking, which results in an AMRI wave located mainly at the top of the cylinder, preferring one of the two possible $m$ modes (refer to \cite{Seilmayer2016a}). 
This is surprising, since the linear theory predicts a non-axisymmetric solution with identical weights for the waves with $m=\pm1$. 

% ------------------------------ Thermal Convection ----------------------------------

\section{The impact of thermal convection\label{sec:EffectOfThermalConvection}}

\begin{table}
	\centering{}
	\caption{\label{tab:Material-Properties}Material Properties GaInSn}
	\begin{tabular}{crl}
		\toprule 
			Property 					& \multicolumn{2}{l}{Value at $\vartheta=\SI{20}{\degreeCelsius}$}\\
		\midrule
		\midrule 
			thermal expansion 		& $\beta_{\mathrm{L}}=$ & \SI{1.24e-4}{\per\kelvin}\\
		\midrule 
			kinematic viscosity 	& $\nu=$ &\SI{0.34e-6}{\square\meter\per\second}\\
		\midrule 
			Prandtl-number 			& $\mathrm{Pr}=$ & \num{0.033}\\
		\midrule 
			thermal diffusivity 	& $\alpha=$ & \SI{10.3e-6}{\square\meter\per\second}\\
		\bottomrule
	\end{tabular}
\end{table}

Our observations indicate that thermal convection is a plausible source of the observed frequency shifting and symmetry breaking. Prior to discussing the details, an estimation of the general impact of thermal convection is considered. 

Table~\ref{tab:Material-Properties} lists the properties of GaInSn, which is the working fluid. GaInSn has a very low viscosity $\nu$ and a rather low thermal diffusivity $\alpha$. Using the properties of GaInSn, the governing Rayleigh number
\begin{equation}
	 \mathit{Ra}=\frac{\beta_{\mathrm{L}}\cdot g\cdot L^{3}\cdot\Delta\vartheta}{\alpha\nu}
	 \label{eq:Ra_norm}
\end{equation}
\noindent ranges from $10^4$ to $10^6$, given a very weak (i.\,e. radial) heat flux of $\dot q \approx \SI{0.1}{\watt\per\square\meter}$ or a minimal temperature difference of about $\Delta \vartheta \approx \SI{0.1}{\kelvin}$ across the cylindrical gap. The expression above can be rewritten into a modified Rayleigh number $\mathit{Ra^*}$ by approximating $\Delta\vartheta \approx \dot{q} \cdot L/\lambda$, which gives
\begin{align}
	       \mathit{Ra^{*}} & =\frac{g\cdot \beta}{\nu\cdot\alpha\cdot\lambda}\cdot\dot{q}\cdot L^{4}\\
                                          & =\frac{g\cdot \beta \cdot \rho \cdot c_{\mathrm{p}}}
					   		   {\nu\cdot\lambda^{2}} \cdot \frac{P}{A} \cdot L^{4} \label{eq:Ra_modified}\\
\log_{10}\mathit{Ra^{*}} & \approx 4\dots6 \nonumber 
\end{align}with $\dot{q}=P/A$ and $\alpha=\lambda/\left(\rho\cdot c_{\mathrm{P}}\right)$. 
Here, the characteristic length $L=h$ corresponds to the cylinder height and $g$ denotes the gravitational constant.
The rather large value of $\mathit{Ra}$ (and hence, $\mathrm{Ra^*}$) indicates a significant tendency for thermal convection to occur even for weak temperature gradients (see Eq. (\ref{eq:Ra_norm})) or heat fluxes (see Eq. (\ref{eq:Ra_modified})).  

Radiation from the central current-carrying copper rod is identified as the major source of heat during the AMRI experiments.
The hot copper surface $A_{1} = \SI{0.0377}{\square\meter}$ of the rod, with $\vartheta_{\mathrm{rod}} < \SI{60}{\degreeCelsius}$, $r_{\mathrm{rod}}=\SI{15}{\milli\meter}$ and an emission coefficient of $\varepsilon_{1} = 0.124$, is insulated by a vacuum, which prevents convective heat exchange. 
Radiation remains and transports some heat radially outward to the cold inner steel surface $A_{2} = \SI{0.0452}{\square\meter}$ of the vacuum insulation with $\vartheta_{\mathrm{VAC}} = \SI{23}{\degreeCelsius}$, $r_{\mathrm{rod}}=\SI{18}{\milli\meter}$ and $\varepsilon_{1} = 0.9$. 
To quantify the amount of radiated heat, we assume the law of heat exchange by radiation
\begin{align}
	\dot{Q} & =\frac{\sigma_{\mathrm{B}}}{\frac{1-\varepsilon_{1}}{\varepsilon_{1}A_{1}}+\frac{1}{A_{1}}+\frac{1-\varepsilon_{2}}{\varepsilon_{2}A_{2}}}	\cdot\left(T_{\mathrm{rod}}^{4}-T_{\mathrm{VAC}}^{4}\right)\label{eq:Rad_Heat}\\
 			 & =\mathcal{O}\left(\SI{1}{\watt}\right),\nonumber 
\end{align}
with $\sigma_{\mathrm{B}}=\SI{5.67e-8}{W.m^{-2}.K^{-4}}$ as Boltzmann constant.  
The temperatures $T$ are in kelvin. Table~\ref{tab:Heat_Flux-Results} shows the average rod temperatures $\vartheta_\mathrm{rod}$ depending on the inflow temperatures $\vartheta_\mathrm{in}$ of the cooling system and in correspondence with central current $I_\mathrm{rod}$.

\begin{table}
	\caption{\label{tab:Heat_Flux-Results}Heat flux as a function of the inflow temperature $\vartheta_\mathrm{in}$ of the coolant surrounding the inner rod.
The heat flux density at the inner cylinder is approx. $\dot{q}\approx P/\left(\SI{0.1}{\square\meter}\right)$.
Absolute values of $P$ are very sensitive to $\varepsilon_{1}$,
whose value is 0.124 here.}
	\begin{centering}
	\begin{tabular}{ccccccc}
		\toprule 
			 &  & \multicolumn{5}{c}{ $\vartheta_{\mathrm{in}}/\si{\degreeCelsius}$ }\\
			\cmidrule{3-7} \cmidrule{4-7} \cmidrule{5-7} \cmidrule{6-7} \cmidrule{7-7} 
			 &  & 20 & 25 &  & 20 & 25\\
			\cmidrule{3-7} \cmidrule{4-7} \cmidrule{5-7} \cmidrule{6-7} \cmidrule{7-7} 
			$I_{\mathrm{rod}}/\si{\kilo\ampere}$ &  & \multicolumn{2}{c}{$\vartheta_{\mathrm{rod}}/\si{\degreeCelsius}$} &  & \multicolumn{2}{c}{$P/\si{\watt}$}\\
		\midrule
		\midrule 
			10 &  & 27.8 & 32.9 &  & 0.12 & 0.27\\
			15 &  & 36.7 & 41.9 &  & 0.38 & 0.55\\
			20 &  & 49.9 & 53.3 &  & 0.82 & 0.94\\
		\bottomrule
	\end{tabular}
\par\end{centering}
\end{table}

Referring to the unifying theory of Grossman and Lohse \cite{Grossmann2000}, the Reynolds number $\mathit{Re}=u_\mathrm{c} \cdot L/\nu$ scales with the Rayleigh number according to
\begin{align}
	\mathit{Re} & \approx0.037\cdot\mathit{Ra}^{1/2}\cdot\mathit{Pr}^{-3/4}\label{eq:Re_Ra_scaling}\\
                            & =\mathcal{O}\left(1000\right) \nonumber
\end{align}
for non-magnetic convective flow regimes of type I (see \cite{Grossmann2000}). The characteristic velocity for the present (non-magnetic) convection problem is about $u_\mathrm{c}=\mathcal{O}(\SI{1}{\milli\meter\per\second})$.
Since AMRI produces velocities in the same range, an interaction of the two effects must be seriously considered.
Later on, the stationary effect of a radial temperature gradient on AMRI is confirmed by utilizing a simulation of the magnetized thermal problem in a cylindrical shell. We also conduct a parameter study that clarifies the $\mathit{Re} \sim \mathit{Ra}$ scaling.

\raggedbottom

% ------------------------------ magnetized thermal convection ---------------------------------

\section{Magnetized thermal convection}

To investigate thermal convection of an electrically conducting fluid exposed to a magnetic field, we first estimate the impact of magnetic field on convective flow in the vertical direction. The momentum equation 
\begin{equation}
	\frac{\partial\vec{u}}{\partial t} + \left( \vec{u} \cdot \nabla \right) \vec{u} = 
		-\frac{\nabla p}{\rho} + 
		\nu\nabla^{2}\vec{u} + 
		\vec{f}_\mathrm{B} + \vec{f}_\mathrm{L}
	\label{eq:Momentum_general}
\end{equation}
\noindent from the set of Navier-Stokes equations includes the buoyancy force and Lorentz force terms.  With $B_\mathrm{z}, B_\mathrm{r}, u_\mathrm z = 0$ the Lorenz force $\vec{f}_{\mathrm{L}} = \frac{1}{\rho} \cdot \vec{J}\times \vec{B}$ is approximately given by
\begin{equation}
	\vec{f}_{\mathrm{L}} \approx -\frac{\sigma}{\rho} 
	                         		u_\mathrm{z} \cdot B_\varphi^2\cdot \vec e_\mathrm{z},
	\label{eq:LorentzForce}
\end{equation}
which is a result of Ohm's law (Eq.~\ref{eq:OhmsLaw}) in the limit $\nabla \phi \approx 0$, (see Sec.~\ref{sec:Simulation} for details). 
This assumption is useful for estimating the thermal steady state of the flow. 
The buoyancy force $\vec{f}_\mathrm{B} = \beta g\partial_r T(r)\vec{\mathrm{e}}_{z}$ originates from a constant heat flux $\dot q$, or temperature gradient $\partial_r T(r)$, as a function of radius. 
When the inner radial boundary surface at $r = r_\mathrm{i}$ heats up, the fluid begins to rise and drives a convective flow along the inner wall with height $h = L$, which is the characteristic length of the system. 
Since the fluid is electrically conducting, the Lorentz force will approximately balance the buoyancy force, i.\,e. $-\vec f_\mathrm B \approx \vec f_\mathrm L$ in the presence of magnetic field. With $r=r_\mathrm o$, $I=\SI{20}{\kilo\ampere}$ and $\Delta T=\SI{0.1}{\kelvin}$
as parameters, it follows that at the outer radius
\begin{equation}
	u_{\mathrm {th}} = \frac{\rho}{\sigma}\beta g \frac{\Delta T(r)}{r_\mathrm o - r_\mathrm i}\left(\frac{2\pi r}{\mu_{0}I}\right)^{2} \approx \SI{0.2}{\milli\meter\per\second} 
	\label{eq:MagVz}
\end{equation}
\noindent is the expression for the characteristic velocity in the axial direction. Since the UDV sensors are located at the outer radius, the magnitude of this characteristic velocity should be recovered during measurement.

\begin{figure}
	\subfloat[\label{fig:ThConMeasSim}Thermal Convection]{\includegraphics[width=0.42\columnwidth]{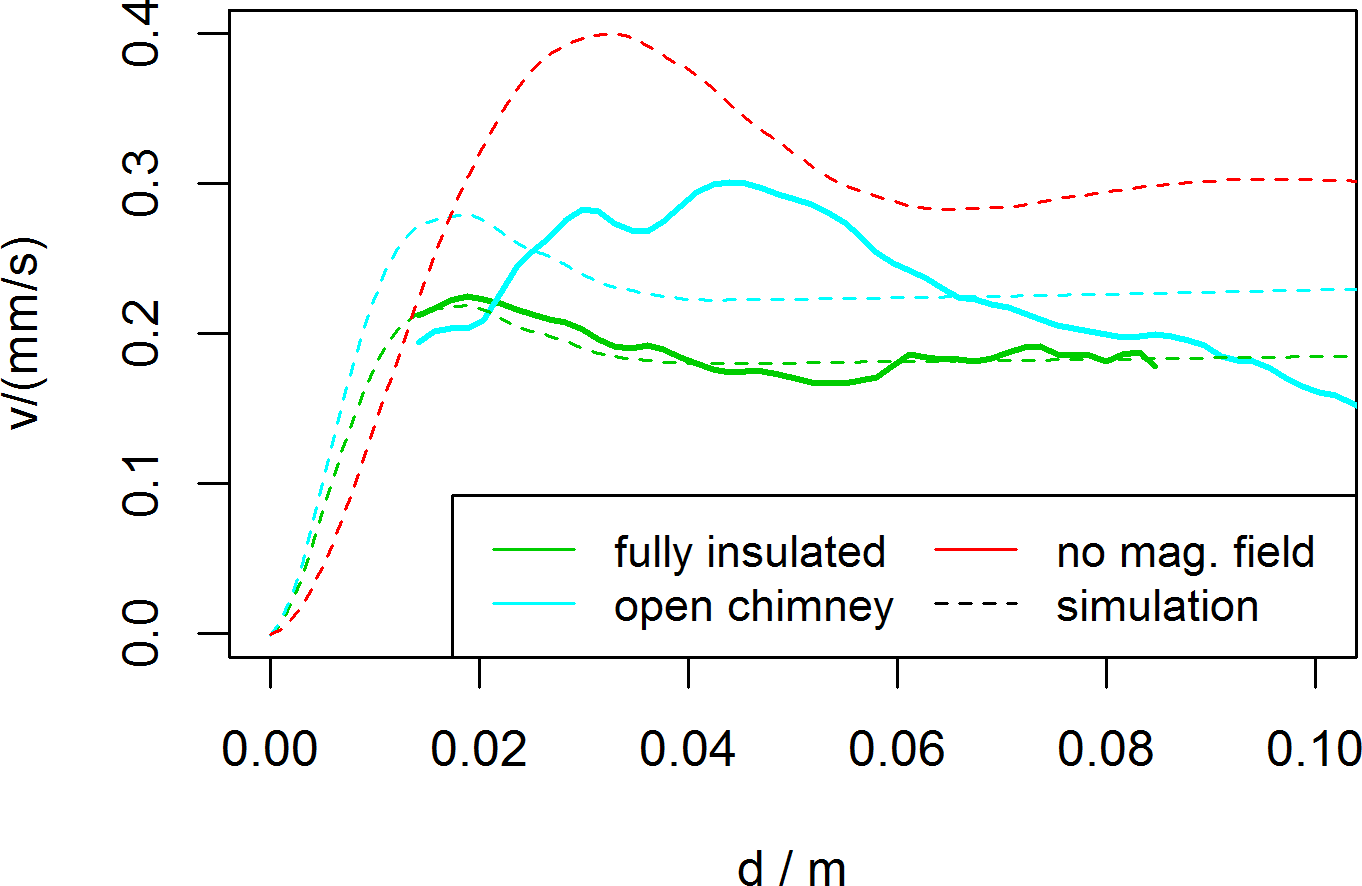}}
		\hfill{}
	\subfloat[\label{fig:Konvektion_Vorzeichenwechsel}Symmetry change]{\includegraphics[width=0.58\columnwidth]{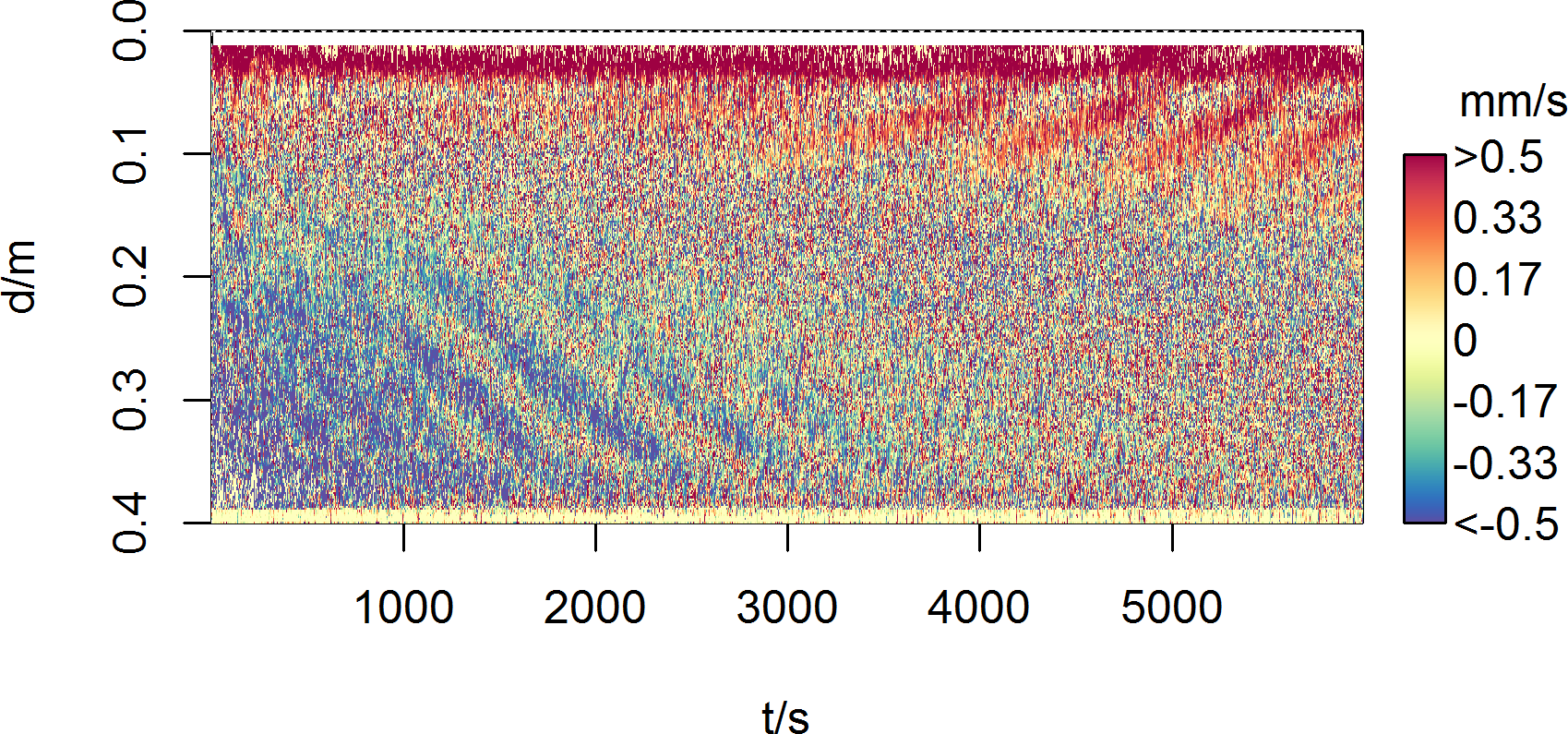}}

	\caption{\label{fig:Thermal-Convection}Thermal convection combined with AMRI.
				a) Thermal convection in the presence of magnetic field for different
				insulation boundaries at $\SI{20}{\kilo\ampere}$. Solid lines show UDV measurements and dashed
				lines correspond to simulations. In case of the ``no mag. field'' simulation, the thermal boundary 
				conditions are taken from the ``fully insulated'' variant, with $B_{\varphi}=0$. 
				b) UDV raw data as 2D time series measured at $\SI{12.87}{\kilo\ampere}\ (Ha=100)$.
				The dominant direction of the AMRI wave depends on the direction of
				thermal heat flux, which points radially inward up to $t =  \SI{3000}{\second}$ and changes then its direction.}

\end{figure}

To clarify the source of thermal heat flux, some simple experiments were conducted. The aim of the experiments was to quantify thermal convection as part of the flow field to which AMRI is superimposed. 
Due to the very low characteristic velocity resulting from the radial temperature gradient, the expected flow structure is assumed to be axisymmetric and stationary. Hence, it is sufficient to leave all rotation for the convection measurement at $\vec{u}_\varphi \approx 0$.
The outer rotation was set to $\Omega_\mathrm o=\SI{1e-3}{\hertz}$ -- a minimal value to capture several complete turns of the cylinder with UDV -- in order to prove its axisymmetry through time- and space-averaged flow patterns. 
Therefore, the zero-velocity condition is only fulfilled approximately. 
Figure~\ref{fig:ThConMeasSim} depicts the result of this velocity measurement. 
Here, the time-averaged velocity profile along the distance $d$ from the sensor is plotted as solid lines. Recall, that $d$ is oppositely directed to the cylinder axis since the UDV sensors point downwards. The magnitude of the determined velocity profile corresponds with the estimate of Eq.~(\ref{eq:MagVz}). 

%------------ Simulation --------------
\section{Simulation\label{sec:Simulation}}

The present axisymmetric problem was investigated via a multi-physics simulation, utilizing COMSOL 5.3a with a rotation-symmetric case study in cylindrical coordinates. 
COMSOL integrates fluid (air and GaInSn) flow, heat conduction and magnetic field influences with the necessary boundary conditions.
The magnetic field simulation was assisted by the `mef' - module, which enables the coupling between Navier-Stokes and Maxwell's equations.
Since the central current-carrying rod is vacuum-insulated, the convective heat transfer by the current is suppressed (see Fig.~\ref{fig:Principle-Sketch}, left).
The suppression of the convective heat transfer results in heat transfer primarily through radiation, as described in section \ref{sec:EffectOfThermalConvection}. 
The inner thermal boundary condition is determined by a constant radial heat flux $\dot q$ of the vacuum insulation (Fig.~\ref{fig:CrossSection} -- (1)) and the inner boundary of the vessel (Fig.~\ref{fig:CrossSection} -- (8)). 
Between the vacuum insulation and the inner boundary of the vessel is a cylindrical gap of \SI{3}{\milli\meter} and \SI{0.4}{\meter} in width and height, respectively.
If the top and bottom ends of the gap are left open, the gap acts like a chimney (see Fig.~\ref{fig:Principle-Sketch} left panel) in conjunction with radiation from the inner rod and an additional heat flux from below. 
Beneath the gap, the base of the current-carrying copper rod heats up the surrounding air, which transports heat into the chimney. 
Thus, the chimney-effect aids and amplifies heat transport within the gap. 
If the top and bottom ends of the gap are closed, as seen on the right hand side of Fig.~\ref{fig:Principle-Sketch}, only internal air circulation transfers heat in the radial direction.
Whichever the case, the inner (cylindrical) liquid metal boundary experiences a radial heat flux (or radial temperature gradient) at any given value of the current, which ultimately drives the vertical convection.

The outer thermal boundary condition was set to that of a free natural convection on a vertical wall. 
In the simulation, the experimental setup was assumed to be a perfectly insulated both from above and below.

\subsection{Definition of electrical potential}
 
Considering the electrical current conservation law and Ohm's law, 
\begin{align}
    \nabla\cdot\vec{J} & = 0 \label{eq:DivJ} \\
				  \vec{J} & = \sigma\left(-\nabla\phi+\vec{u}\times\vec{B}\right) \label{eq:OhmsLaw}
\end{align}
we must now take the electric potential $\phi$ into account, in contrast to the approximation from equation (\ref{eq:LorentzForce}). By combining the two equations (\ref{eq:DivJ}) and (\ref{eq:OhmsLaw}), we find $\nabla\cdot\left(-\nabla\phi+\vec{u}\times\vec{B}\right) = 0$, which can be rewritten as
\begin{equation}
 \laplace \phi =\vec{B}\cdot\left(\nabla\times\vec{u}\right)+\vec{u}\cdot\left(\nabla\times\vec{B}\right). \label{eq:ElPot}
\end{equation}
Note that equation~(\ref{eq:ElPot}) depends on two different components, $\nabla\times\vec{u}$ and $\nabla\times\vec{B}$, which are measures of the contributions of velocity and current to the electric potential.
The COMSOL simulation takes this into account by utilizing Ohm's law within the mef-module.

\subsection{Simulation Results and Discussion}

\begin{figure}
	\subfloat[\label{fig:Principle-Sketch}Principle Sketch]{
		\raisebox{-0.5\height}{\includegraphics[width=0.49\columnwidth]{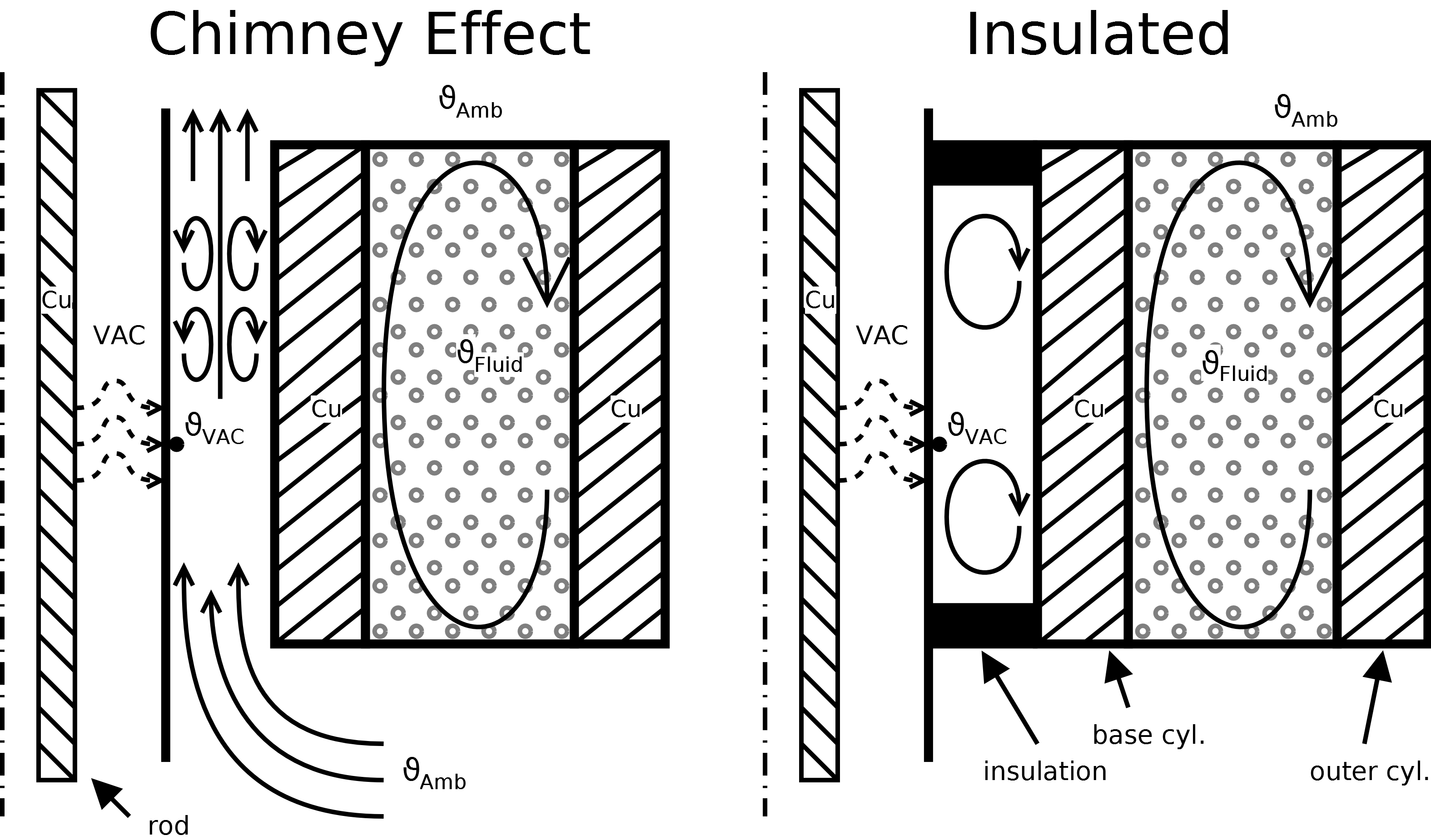}}
	}
		\hfill{}
	\subfloat[\label{fig:ParameterStudy}Parameter Study]{
		\raisebox{-0.5\height}{\includegraphics[width=0.49\columnwidth]{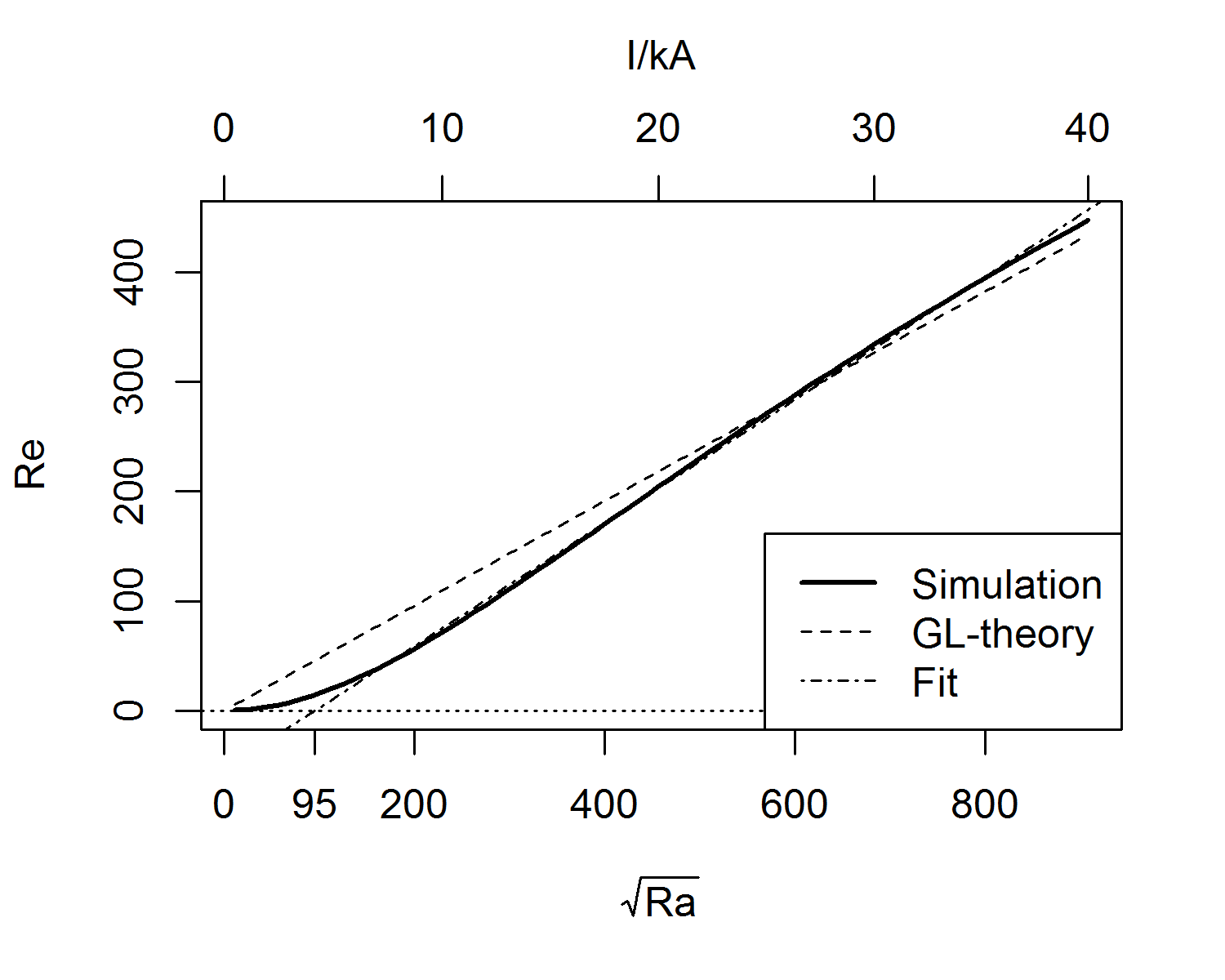}}
	}

	\caption{\label{fig:Thermal-Simulation}Simulation conditions and results.
				a) Sketch showing the cross-section of the TC cylinder from Fig.~\ref{fig:Experimental-setup} to explain the origin of thermal convection. 
				From left to right of Figure~\ref{fig:CrossSection}:
				copper rod (3); vacuum insulation (1); air gap; center cylinder (8)
				and inner cylinder (7); fluid; outer cylinder (5). 
				b) Parameter study $\mathit{Re} \sim \sqrt{\mathit{Ra}}$ is given. The dash-dotted line gives the linear fit starting from $\sqrt{\mathit{Ra}}>250$.
				} 

\end{figure}

Figure~\ref{fig:Principle-Sketch} shows the two major convective flow mechanisms with respect to the \emph{technical} thermal boundary conditions. Figure~\ref{fig:ThConMeasSim} is a summary of the convection measurements of the axisymmetric $u_\mathrm z$ component (solid lines), which occur for a hydrodynamically stable TC configuration with $\Omega_\mathrm i = 0$, $\Omega_\mathrm o \approx 0$ and $\SI{20}{\kilo\ampere}$, with the convection in the flow being driven by the radiation from the central rod. 
The experimental results of the different insulation schemes (open chimney and insulated) lend credence to the fact that the proposed chimney-effect is present and enhances heat transport.
The dashed lines in Figure~\ref{fig:ThConMeasSim} depict the results of the simulation to verify the convective flow pattern that were obtained experimentally. 
As we can see, the simulated result is in good agreement with the main flow features of the insulated setup. For the open chimney configuration, the maximum velocity of the simulated result corresponds to that of the experiment, but the locations of the maximum velocity differ. For the closed chimney (insulated) case, the parameter study in Figure~\ref{fig:ParameterStudy} confirms the relation $u_\mathrm {th}:=f(I^2)$. The result of the parameter study is the relation $\mathit{Re} = 0.043\cdot \mathit{Ra}^{0.5} \cdot  \mathit{Pr}^{-0.75}$, further confirming the $\mathit{Re}\sim \mathit{Ra}^{0.5}$ scaling for convective flow according to Grossman and Lohse \cite{Grossmann2000}. However, the value of the coefficient, which is \num{0.043} here, differs from the standard non-magnetic Rayleigh-B\'ernard case. A nonlinear onset region which ends at $\sqrt{\mathit{Ra}} \approx 95$ was also identified. Studies on vertical convection in a cylinder conducted by Shishkina et al. \cite{Shishkina2016} additionally support the scaling law found here.

%------- Experimental Findings and Discussion -----%
\section{Discussion}

An intuitive analogy of the underlying mechanism by which convection affects AMRI is that weak axisymmetric thermal convection acts like conveyor belt with velocity $u_\mathrm{th}(r,I)\cdot \vec{\mathrm e}_z$, which transports the phase velocity of AMRI so that $v = v_\mathrm{AMRI} - u_\mathrm{th}(I)$.
As a result, the AMRI wave propagates in a moving frame of reference leading to its fractionally shifted characteristic frequency $\omega = (v_\mathrm{AMRI} - u_\mathrm{th})\cdot k$.
For a non-magnetized wide-gap Taylor-Couette flow with radial heat flux, similar effects were discovered by Guillerm et al. \cite{Guillerm2015}. 
In their experimental and numerical work, it was shown that in a marginally unstable hydrodynamic flow with a weak thermal convection relative to the angular velocity of the system, i.\,e.\ $u_\varphi \gg u_\mathrm{th}$, the thermal convection lifts the stationary Taylor vortices, so that the initially stationary Taylor vortex flow (TVF) becomes an axisymmetric traveling wave. 
The interpretation of this effect is that the resulting phase velocity $v = v_\mathrm{TVF} + u_\mathrm{th}$ is a linear superposition of the TVF and thermal convection.

With regard to the results in Fig.~\ref{fig:AMRI-results}, the systematic deviation from the linear theory is caused by the weak thermal convection. The dashed lines indicate the shifted predictions under the assumption of the dependence of $u_\mathrm{th}$ on current and a reduced critical current of $I_\mathrm{crit} = \SI{7.8}{\kilo\ampere}$, which follows from Fig.~\ref{fig:Energy}. Since this point marks the onset of instability, data points regarding the drift rate or wave number below the critical current are not related to AMRI.
The latter is a critical assumption since there is no clear theoretical proof pertaining to AMRI. 
However, several related works, which show that radial thermal heat flux shifts critical parameters and create instability, exist. 
The works by Takhar et al. \cite{Takhar1988} and Guillerm et al. \cite{Guillerm2015} (with a few restrictions) showed that the critical rotation rate required to destabilize a non-magnetized TC system (even for second order instabilities) is increased by radial heat flux at the inner cylinder. 
For the  magnetized case, Rüdiger et al. \cite{Rudiger2009} come to the conclusion that in the presence of density stratification, the necessary critical magnetic field increases. 
We might guess that for the non-stratified case as ours, the opposite effect, i.\,e., the lowering of the critical magnetic fields, takes place.
This argument is strengthened by the findings of Takhar et al. \cite{Takhar1992}, which pointed out that in a wide gap TC setup with $r_\mathrm i / r_\mathrm o = 0.5$, axial magnetic field $B_\mathrm z$ and a radial temperature gradient, the critical velocity increases with heat flux.
With respect to the $\mathit{Re} \sim \mathit{Ha}$ parameter space from \cite[Fig. 1b]{Hollerbach2010}, the entire curve of marginal instability would be shifted upwards.
Provided that the Reynolds number remains constant, this shift would lower the critical current. However, characteristic frequencies can only be adjusted properly if the shift of the critical current by $\Delta I \approx \SI{2.6}{\kilo\ampere}$ is not disregarded.

In addition to the findings above, we observe a symmetry breaking that depends on the direction of heat flux. Figure~\ref{fig:Konvektion_Vorzeichenwechsel} shows the time evolution of AMRI wave produced when $\Omega_\mathrm{o} / \Omega_\mathrm{i} = \num{0.26}$ and $\SI{12.87}{\kilo\ampere}\ (\mathit{Ha}=100)$, where the thermal gradient reverses during the observed period. 
Initially, the current-heated yellow coil (see Fig.~\ref{fig:CrossSection}) provides a negative radial heat flux (heat flux points radially inwards) with the temperature of the fluid initially at $\vartheta_\mathrm{Fluid}\approx \SI{30}{\degreeCelsius}$. 
When the coil current is switched off at $t=0$, the coil begins to cool.
At first, the AMRI wave -- modified by the upward convection at the outer cylinder -- is concentrated in the lower half of the volume, which is contrary to normal behavior (refer to \cite{Seilmayer2016a}). 
Due to the cooling of the coil, the heat flux points radially outwards again at $t \approx \SI{3000}{\second}$. 

When the reversal of the heat flux occurs, the wave moves from the lower to the upper half of the volume, then continues to evolve.
Finally, it must be emphasized again that the PROMISE experiment was not designed for thermodynamic experiments. 
Therefore, precise measurements of the wall temperatures of the rotating cylinders, especially of the inner cylinder, were not possible.

%---- Conclusion ---

\section{Conclusion}

The primary findings of this paper are the symmetry breaking of the AMRI wave and the systematic shift of its characteristic frequencies due to thermal boundary conditions. Thermal convection driven by radiative heat flux from the central current-carrying rod was identified as the main cause.
The analysis of (non-magnetized) thermal convection according to Grossmann and Lohse~\cite{Grossmann2000} predicts $v_\mathrm z = \mathcal O(\SI{1}{\milli\meter\per\second})$, which is higher than the measured characteristic velocity in the magnetized case. This difference stems from the fact that the Lorentz force $f_\mathrm L$ decelerates the rising fluid, thereby reducing the convection within the fluid. Experiments and simulations give clear evidence that the two types of insulation, open chimney and fully insulated, lead to different amplitudes of convection. The simulated results of the closed chimney configuration fits better with its experimental counterpart. The simulation results indicates that $\mathit{Re} \sim \mathit{Ra}$, which is in accordance with the frequency shift of the developing traveling wave.
 
When the radial direction of the heat flux was changed by heating the experimental setup from the outside, a mode change from $m=1\ \rightarrow\ m=-1$ was observed for $I = \SI{16}{\kilo\ampere}$ (see Fig.~\ref{fig:Konvektion_Vorzeichenwechsel}). 
The question remains as to whether the shift in the characteristic frequencies of AMRI can be explained entirely by the (non-) linear superposition with thermal convection. 
The early onset of the instability (see Fig.~\ref{fig:Energy}) at \SI{7.8}{\kilo\ampere} may be explained by the numerical analyses by Takhar et al.~\cite{Takhar1992}, which shows that thermal convection shifts the entire parameter space to that corresponding to higher critical Reynolds numbers. 
Since the analyses were for cases with axially applied magnetic field, further detailed theoretical investigation must be conducted for the azimuthal magnetic field-induced AMRI.

% ------------------------------------ Acck --------------------------------
\section*{Acknowledgements}

This project has received funding from the European Research Council (ERC) under the European Union's Horizon 2020 research and innovation program (grant agreement No 787544). The authors would like to thank Kevin Bauch for carrying out many of the experiments.

% ------------------------------------ Bib --------------------------------

\bibliographystyle{mhd}
\bibliography{ThermalAMRI}

\lastpageno
\end{document}